\documentclass[10pt,letterpaper]{article}
\usepackage{opex3}

\usepackage{cite}
\usepackage{amsmath}    
\usepackage{graphicx}   
\usepackage[dvipdfx,cmyk]{xcolor}     
\usepackage[caption=false]{subfig}  
\usepackage[colorlinks,linkcolor=red,citecolor=brown,urlcolor=blue]{hyperref}
\usepackage{dsfont,bm}
\usepackage{color}

\newcommand{\be}{\begin{equation}}
\newcommand{\ee}{\end{equation}}
\newcommand{\bA}{\begin{align}}
\newcommand{\eA}{\end{align}}
\newcommand{\bF}{\begin{figure}}
\newcommand{\eF}{\end{figure}}

\begin{document}
\title{Strain-optic active control for quantum integrated photonics}
\author{Peter C. Humphreys,\textsuperscript{1} Benjamin J. Metcalf,\textsuperscript{1} Justin B. Spring,\textsuperscript{1} Merritt Moore,\textsuperscript{1} Patrick S. Salter,\textsuperscript{2} Martin J. Booth,\textsuperscript{2} W. Steven Kolthammer\textsuperscript{1} and Ian A. Walmsley\textsuperscript{1,*}}
\address{\textsuperscript{1}Clarendon Laboratory, Department of Physics, University of Oxford, OX1 3PU, United Kingdom\\ \textsuperscript{2}Department of Engineering Science, University of Oxford, Parks Road, Oxford, OX1 3PJ UK}
\email{\textsuperscript{*}i.walmsley1@physics.ox.ac.uk} 

\begin{abstract}

We present a practical method for active phase control on a photonic chip that has immediate applications in quantum photonics. Our approach uses strain-optic modification of the refractive index of individual waveguides, effected by a millimeter-scale mechanical actuator. The resulting phase change of propagating optical fields is rapid and polarization-dependent, enabling quantum applications that require active control and polarization encoding. We demonstrate strain-optic control of non-classical states of light in silica, showing the generation of 2-photon polarisation N00N states by manipulating Hong-Ou-Mandel interference. We also demonstrate switching times of a few microseconds, which are sufficient for silica-based feed-forward control of photonic quantum states.
\end{abstract}


\bibliographystyle{osajnl}
\bibliography{stressPapers.bib}

\section{Introduction}

Linear optics provides a valuable platform for quantum information processing. Current approaches to implementing all-optical computing schemes have focused on the use of guided waves, harnessing advances in photonic technology that have accompanied the emergence of optical communications platforms~\cite{Thompson2011}. For these applications, two important practical requirements are total device efficiency and, for proof-of-concept work in particular, ease of fabrication. These issues have put silica-based devices at the forefront of integrated QIP experiments. Silica photonic chips provide low propagation losses, efficient coupling to silica fiber - providing a robust and convenient ultra-low loss delay line, and fabrication by direct laser writing~\cite{DellaValle2009b}. Silica waveguides have been used as quantum light sources \cite{Spring2013, Fiorentino2002}, quantum channels connecting remote nodes~\cite{Marcikic2003} and integrated optical circuits as a means to generate the interference needed for computational protocols~\cite{Politi2008, Spring2012}. 

An outstanding technical requirement for optical quantum information processing is low-loss, rapid active control of spatial and polarisation modes. This is necessary for a host of essential applications including multiplexing heralded light sources~\cite{Ma2011, Meany2014, Christ2012}, feed-forward control for linear-optical quantum computing~\cite{Kok2007} and quantum communication~\cite{Marcikic2003}. Fast spatial and polarisation control with quantum light has been shown using the electro-optic effect in lithium niobate~\cite{Bonneau2012} but relatively inefficient coupling to silica fibers~\cite{Alibart2005, Alferness1982} and increased complexity of fabrication currently present roadblocks to near-term quantum application. For standard silica devices the electro-optic effect is non-existent, and instead, experiments to date have relied on thermo-optic control~\cite{Smith2009}. This approach, however, offers no polarization sensitivity, is inherently slow (on the ms timescale), and is fundamentally incompatible with cryogenic detectors due to the associated heat load. It is also incompatible with typical femtosecond-laser written devices~\cite{DellaValle2009b} due to the thickness of typical substrates, leaving these devices with no mechanism for active control. Another approach recently demonstrated in silicon nitride uses nano-electromechanical structures~\cite{Poot2014} to achieve sub-$\mu$s switching speeds. However, this requires relatively sophisticated fabrication and has not been implemented in silica.

\section{Experimental overview}

Here we show strain-optic control to be an attractive method for active reconfiguration of quantum photonic circuits. The essential idea is that a stress is applied to the region in which light propagates, creating a local change in the refractive index and an corresponding phase shift for the guided waves. This effect has previously been employed to create polarisation independent phase shifters on silica-on-silicon telecommunications devices~\cite{Donati1998}. We extend this idea, capitalising on the fact that strain generally causes anisotropic changes in refractive index, and thus results in birefringence aligned to the strain field~\cite{Tsia2008a}. Our demonstration uses a millimeter-scale actuator to achieve a polarization-sensitive effect that may be readily applied to common platforms, including direct laser-written waveguides in silica. Operation is shown on the microsecond timescale, sufficient to implement feed-forward control entirely within silica guided-wave devices. The switching method introduces no excess loss beyond the propagation loss associated with the silica substrate, allowing the realisation of low-loss polarisation switches. This approach is a promising near-term route to achieving scalable reconfigurable quantum information processing channels, based on a combination of polarisation, spatial and temporal coding~\cite{Humphreys2013a}. 
\begin{figure}[htbp]
\begin{center}
\includegraphics[width = 4.5cm]{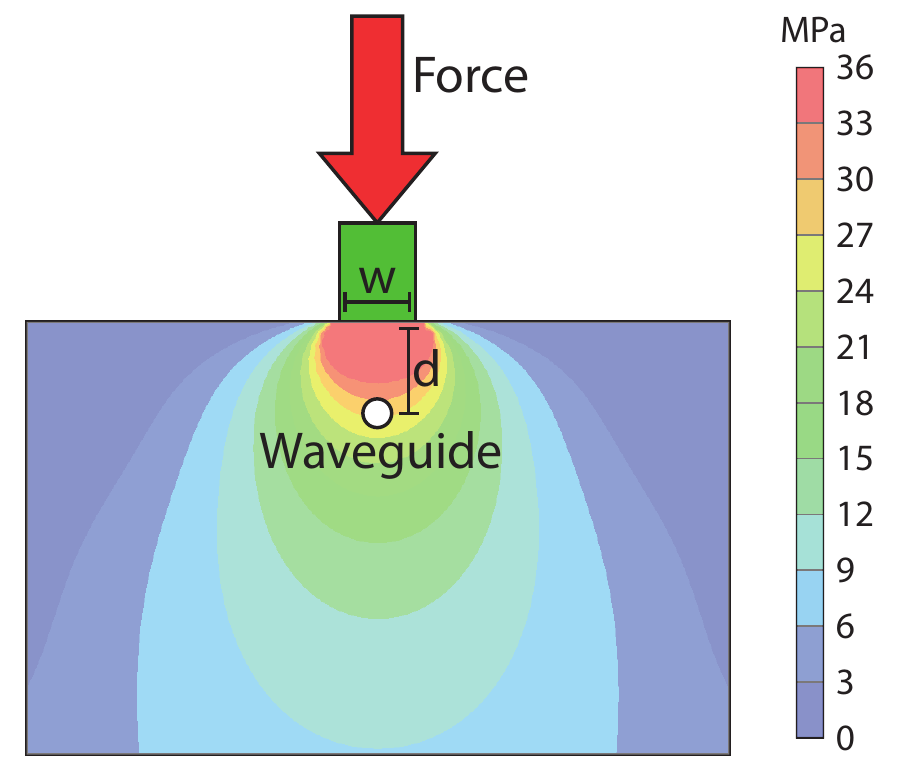}
\caption{Cross section of the strain-optic phase shifter. A force (red arrow) applied to a steel ram (green box) by a charged piezoelectric element modifies the stress (theoretical contours shown) at the location of a waveguide (white circle).}
\label{fig:stressor}
\end{center}
\end{figure}

Our device, illustrated in Fig.~\ref{fig:stressor}, is described by a simple model using the well-known photoelastic effect. An external pressure is applied over a region of the chip's surface, resulting in an elastic stress, and corresponding strain, throughout the chip. At the location of a buried waveguide, the change in the refractive index $n$ along direction $i$, where $i \in \{x,y,z\}$, is given by $\delta n_i = \sum_j -\frac{1}{2} n^3 \rho_{ij} \sigma_j / E$, where $E$ is the Young's modulus, $\sigma_j$ is the stress along direction $j$, and $\rho$ is the (polarisation dependent) strain-optic coefficient tensor. Assuming that the ram is centered over the waveguide, and that the stress is therefore vertically aligned (direction $z$), allows us to simplify this to $\delta n_i = -\frac{1}{2} n^3 \rho_{i z} \, \sigma_z / E$. The result is a change in the propagation constant of a guided mode compared to an unstrained guide. For silica, $\rho_{xz} = \rho_{yz} = 0.12$ and $\rho_{zz} = 0.26$~\cite{Bertholds1988}; it is this difference that creates a birefringent effect. 

The stress required to induce a given phase shift $\theta$ is a function of the length $l$ of the stressed region, and is approximated by
\begin{equation}
\sigma_z = \frac{E \theta  \lambda }{\pi  l n^3 \rho_{i z}}.
\end{equation}
For example, in order to induce a phase shift of $2\pi$ for vertically polarised 830 nm light in a silica waveguide with a stressed region of length 1 mm, a change in refractive index of $\delta n = 8.3 \times 10^-5 $ must be induced (much smaller than the waveguide index contrast of $\delta n \approx 5 \times 10^-3$). This corresponds to a stress of 14 MPa, roughly two orders of magnitude less than the compressive strength of silica.

We demonstrated this phase shifting method using direct-written waveguides in a fused-silica photonic chip (Lithosil Q1). These waveguides were fabricated using a regeneratively amplified Ti:Sapphire source delivering 100~fs pulses at a repetition rate of 1~kHz. The beam was focused with a 0.5~NA, 20$\times$ objective, and the fused silica substrate translated through the focus at 25~$\mu$m/s perpendicular to the optic axis. Slit beam shaping was applied to control the structural cross-section of the waveguide~\cite{Salter2011}. The fabricated waveguides displayed typical values for propagation loss \textless  0.5 dB/cm, coupling loss to single mode fiber \textless 0.5 dB and birefringence \textless  2$\times 10^{-5}$~\cite{Spring2014}.

We generate an external pressure on the chip surface by pressing a steel ram with a contact surface 1 mm long by 0.1 mm wide onto the surface of the silica above a waveguide. The force on the ram is generated either by turning a screw in contact with its upper surface or by charging a piezoelectric element positioned between the screw and ram. A force of only 20 N is needed to create a stress of 14 MPa under the ram; this is readily generated by the 3x3x2 mm piezoelectric element (Physik Instrumente PICMA Chip Actuator) that we employed. As shown in Fig.~\ref{fig:stressor}, the stress at a waveguide depends on its depth. Unless specified otherwise, we carry out our experiments on waveguides located $100 \, \rm{\mu m}$ from the silica surface, resulting in a stress approximately 80\% of that immediately under the ram.

\section{Classical operation}

\begin{figure}[htbp]
\begin{center}
\includegraphics[width = 9cm]{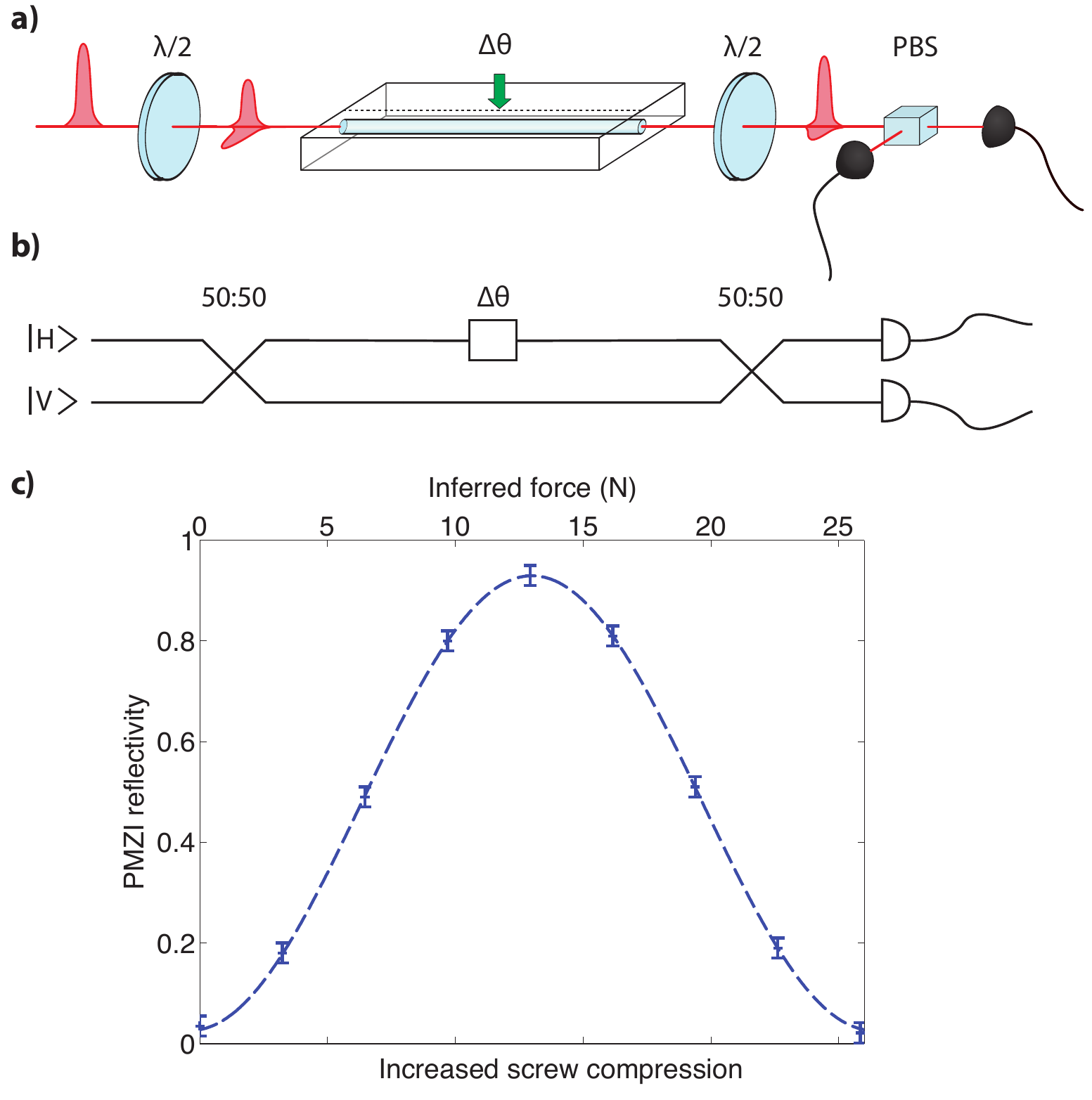}
\caption{a) Schematic of the polarisation Mach-Zehnder interferometer built using the strain-optic device sandwiched by two half-wave plates. Applying stress to the waveguide induces a birefringence, leading to a relative phase shift between two polarisation components in the guide. b) Equivalent mode representation where each polarization is depicted as a spatial mode. c) The normalised intensity of one polarisation output from the PMZI as a function of increased screw compression. The inferred force was estimated from the birefringence that would be required to produce the measured change in PMZI reflectivity. Sinusoidal modulation is observed, with a fringe visibility of $97\pm3\%$.}
\label{PMZI}
\end{center}
\end{figure}

To construct a polarisation switch, we embed the manipulated waveguide between static waveplates, used to rotate the polarisation axes of the light by $\pi/4$. We achieve this with waveplates external to the chip, but these elements may alternatively be built directly in guiding structures using static stresses \cite{Rechtsman2012,Fernandes2012,Corrielli}. The two waveplates create a polarisation Mach-Zehnder interferometer (PMZI), as shown in Fig.~\ref{PMZI}(a,b). A strain-induced modification to the waveguide, such that a phase shift of $\pi$ radians is created in one polarisation relative to the other, will switch the polarisation of the modes at the output of the PMZI. The relative birefringent phase shift, and therefore the change in the PMZI output intensity $I$ induced by a stress $\sigma_z$, can be determined from 
\begin{equation}
I = \cos^2(\Delta \theta/2) \text{, where } \Delta \theta = \pi n^3 (\rho_{z z} - \rho_{x z}) \, \sigma_z l / (E \lambda)\label{eqn:fringe}.
\end{equation}

Control of the polarization of classical light with a PMZI is shown in Fig.~\ref{PMZI}(c). The power at the output ports was recorded at certain screw positions as the screw compression was increased by hand. The interference visibility was determined by manually positioning the screw to maximize and minimize the reflectivities. While absolute calibration of the reflectivity determined by a given screw position is not feasible due to trial-to-trial variation in mechanical alignment, the estimated force on the ram shown was calculated from the observed reflectivity and Eq.~\ref{eqn:fringe}. With the piezo-electric element added between the screw and ram, as described above, 60 V corresponded to a $\pi$ phase shift, and arbitrary reflectivities were reproducible within 1\%. With both drive methods, reflectivity drifts of maximally 10\% occur over a few hours of constant application. We suspect that this can be rectified through improvements to the mechanical rigidity in the mounting of the piezoelectric above the waveguide.

\begin{figure*}
\begin{center}
\includegraphics[width = 13cm]{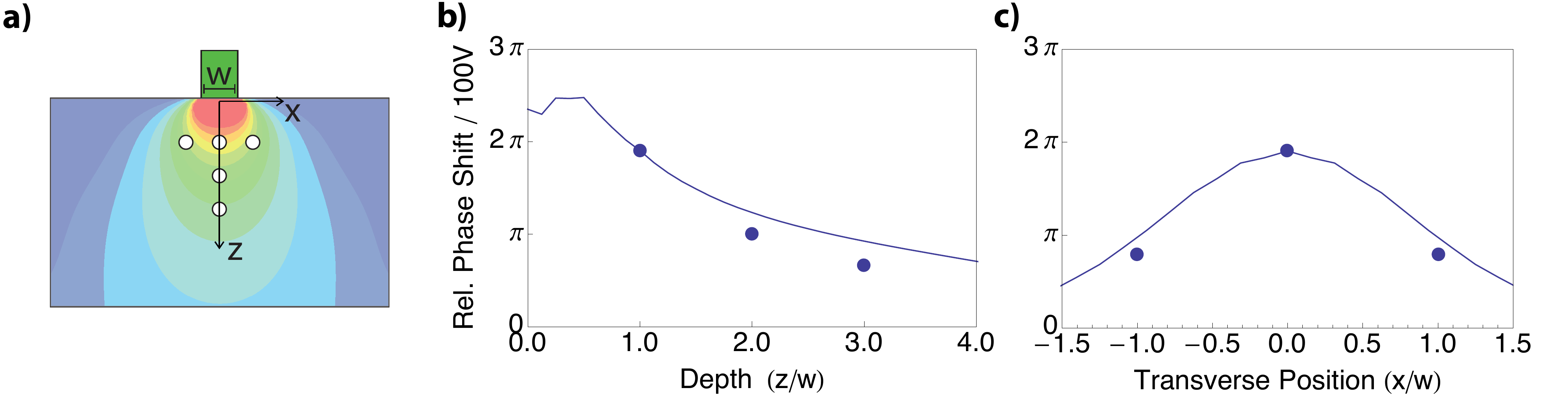}
\caption{a) An array of waveguides (white) was used to map out the relative phase shift induced between vertical and horizontal polarisations in the region of the stressor. Shown are relative phase shifts as a function of b) depth and c) transverse position, in units of the ram width $w = 100 \mu m$. The transverse measurements are from waveguides at depth $w$. The solid line shows data from a finite-element calculation, scaled to match the data point at depth $w$ and zero transverse position.}
\label{fig:mappedOutStress}
\end{center}
\end{figure*}

The profile and extent of the stress field (Fig.~\ref{fig:stressor}) determines the effect of a strain-optic controller on neighboring waveguides, an important issue for multi-path integrated architectures. To test the theoretical model, we measure the induced relative phase shift for an array of guides written at different depths and horizontal positions relative to the stressing element, as illustrated in Fig.~\ref{fig:mappedOutStress}(a). The resulting relative phase shifts are shown in Fig.~\ref{fig:mappedOutStress}(b) and Fig.~\ref{fig:mappedOutStress}(c). As can be seen, the strain field transverse decay length (to 10\% of its initial value) is $2 \, w$, where $w$ is the ram width, suggesting that the effect can be readily localised. 

\section{Non-classical Hong-Ou-Mandel interference}

\begin{figure}[htbp]
\begin{center}
\includegraphics[width = 6.5cm]{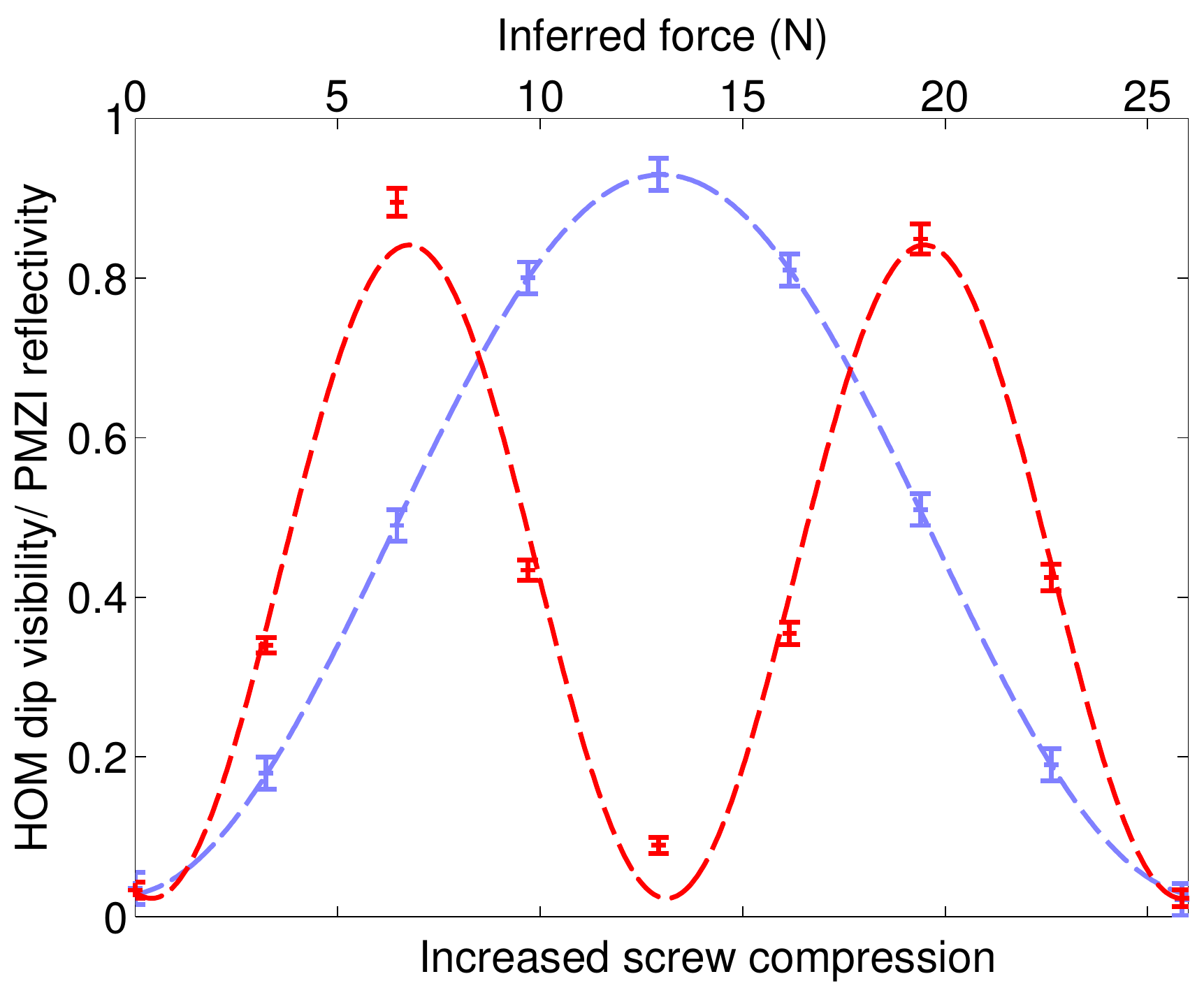}
\caption{Quantum Hong-Ou-Mandel interference between two single photons input to the PMZI. A modulation in the interference visibility (red) is seen as a function of the screw compression. For reference the classical interference (Fig.\ref{PMZI}) is also shown (blue).}
\label{fig:PMZIfringesQuantum}
\end{center}
\end{figure}

Importantly, since for our stressor design $\delta n \ll n$, there is no scattering due to the induced strain, so that the overall transmission of the device is unaffected by including the switching element. This is borne out by our experiments, in which a variation in transmission of less than 1\% is seen. This low loss makes the strain-optic control method particularly well suited to quantum applications. We demonstrated this by inputting heralded single photons from two spontaneous parametric down-conversion (SPDC) sources to the PMZI, with one photon in each polarisation. At a fixed stress level, scanning the temporal delay between the photons produced a non-classical Hong-Ou-Mandel interference dip \cite{Hong1987} in the photon detection coincidences. Setting the temporal delay between the photons to zero maximises the photon indistinguishability. At this setting, applying stress modulates the visibility of the interference with a period half that of the classical fringes.  The enhanced modulation, shown in Fig. \ref{fig:PMZIfringesQuantum}, is a feature of the two-photon N00N state~\cite{Thomas-Peter2011}.

\section{Switch time}

\begin{figure}[htbp]
\begin{center}
\advance\leftskip-0.25cm
\includegraphics[width = 8cm]{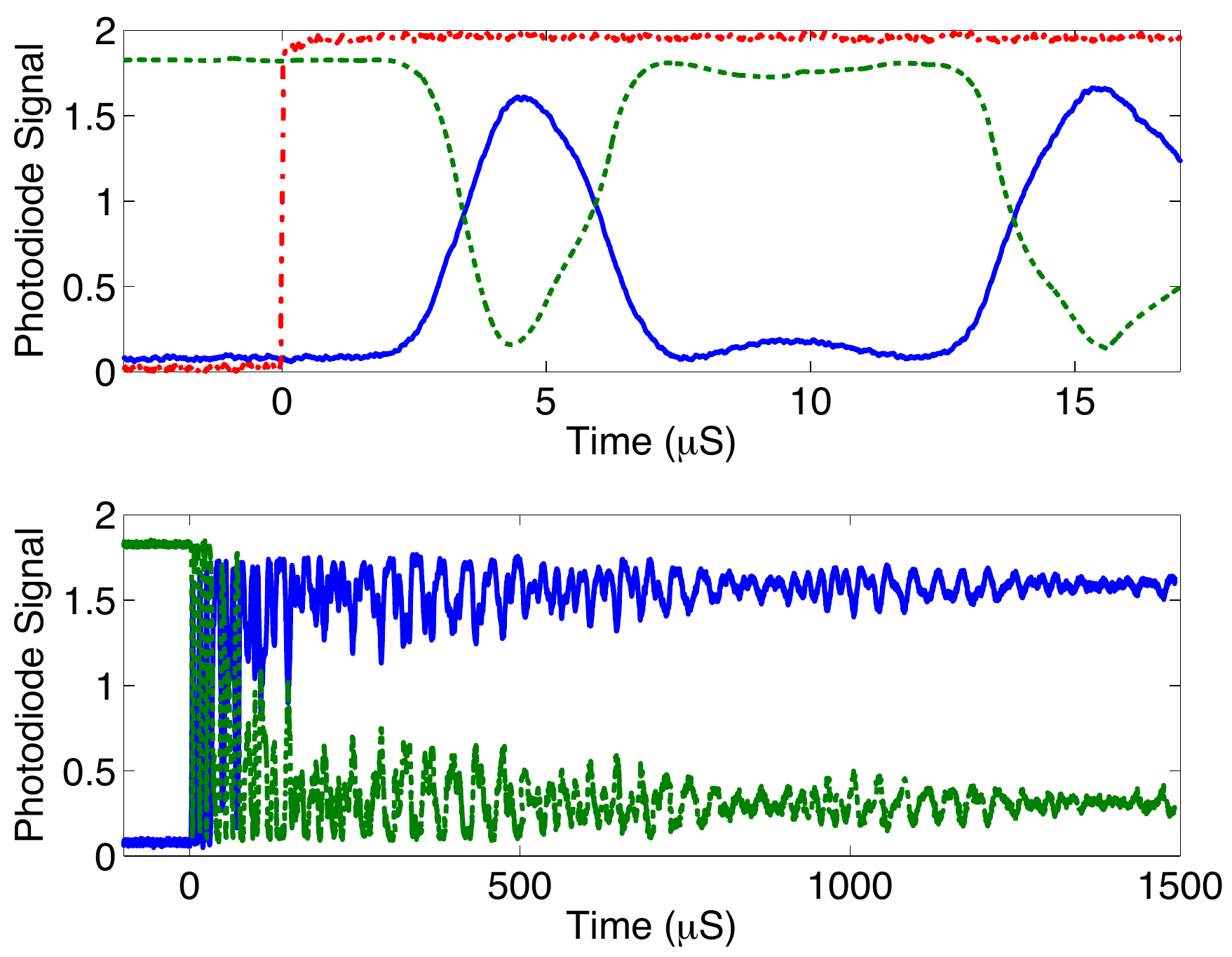}
\caption{Demonstration of the switch time. A control signal (red dot-dashed line) initates the switch operation, leading to the discharge of a piezoelectric actuator. The transmission of each polarisation component (blue solid line and green dashed line) through the PMZI shows (a) a rise time of 1.7 \,$\mu$s (with subsequent mechanical ringing) and (b) a reset time of 1.5 ms due to this ringing.}
\label{fig:FastPiezo}
\end{center}
\end{figure}
To investigate switching time, we apply a current pulse to the piezoelectric actuator clamped above the ram. A single field-effect-transitor switch is gated by a low-voltage square wave from a waveform generator to apply up to 70\,V across the actuator. An example of the PMZI output response is shown in Fig.~\ref{fig:FastPiezo}. The switch response comprises a trigger-to-switch delay time of 2 $\mu$s and rise time (10\% to 90\%) of 1.7 $\mu$s. The observed switch time is limited by the fundamental resonance of the coupled piezoelectric-silica system, due to undesired mechanical oscillations. In this simple configuration, the reset time for the switch is approximately 1\,ms because of the reflection of the acoustic wave induced by the pulsed strain at the chip surface. This can be mitigated by appropriate impedance matching into a damping material, such as aluminum~\cite{Wang1990}. The practical consequence of a non-negligible reset time is a reduction in the repetition rate for a quantum optics experiment or the effective clock rate of an optical information processor.

Even with our un-optimized design, the observed switch time is sufficient for immediate use in proof-of-concept quantum information applications involving feed-forward control. A few-microsecond optical delay can be achieved with a sub-km fiber path, which introduces less than 0.2 dB loss (in addition to the pre-existing \textless 0.5 dB coupling loss). We note that the ultimate limit of the switch rise time is determined by the velocity of sound, $v$, in the material. A simple estimate is given by the time taken for a strain wave to propagate across the guided mode; for $v = 3000$ m/s and a mode diameter of 10\,$\mu$m, this is 5\,ns. This limit could be achieved with an impulse ram that strikes the chip surface, which is contrast to our case in which the ram and chip begin in contact.

\section{Conclusion}

In conclusion, we have demonstrated a method for polarisation-dependent active phase control in silica waveguide photonic chips based on the strain-optic effect. This approach is inherently low-loss, making it ideally suited to the stringent requirements of quantum optical applications. We demonstrated this by using this switching method to control the on-chip quantum interference of two single photons. Further, we demonstrated that the technique allows rapid switching, which opens the door to new applications in feed-forward optical quantum control.

\section*{Acknowledgments }We thank T Hiemstra, XM Jin and M Barbieri for helpful discussions. This work was supported by the EPSRC(EP/K034480/1, EP/H03031X/1, EP/H000178/1, EP/E055818/1), the EC Marie Curie ITN PICQUE, and the AFOSR EOARD. WSK is supported by EC Marie Curie fellowship (PIEF-GA-2012-331859).

\end{document}